\documentclass[prb,reprint,citeautoscript]{revtex4-1}

\usepackage{amsmath}
\usepackage{amssymb}
\usepackage{graphicx}
\usepackage{hyperref}
\newcommand{\overbar}[1]{\mkern 4mu\overline{\mkern-4mu#1\mkern-4mu}\mkern 4mu}
\begin{document}
\title{Statistics of radiation due to non-degenerate Josephson parametric down conversion}

\author{Lisa Arndt}
\email{lisa.arndt@rwth-aachen.de}
\affiliation{JARA Institute for Quantum Information, RWTH Aachen University, 52056 Aachen, Germany}

\author{Fabian Hassler}
\affiliation{JARA Institute for Quantum Information, RWTH Aachen University, 52056 Aachen, Germany}

\date{July 9, 2019}
\begin{abstract}
In a process called parametric down-conversion, a dc-biased Josephson junction coupled to two microwave resonators emits photon pairs when the Josephson frequency matches the sum of the two resonance frequencies. Recent experiments have shown that such a setup permits analyzing the correlation of the radiation. Motivated by these results, we study theoretically the full counting statistics of a non-degenerate parametric oscillator below the threshold of instability. Furthermore, we analyze the second-order coherences of the radiation and discuss thermal effects on the radiation statistics. We provide results for the driving strength at which the Cauchy-Schwarz inequality is most strongly violated. Additionally, we study the impact of asymmetry in the linewidth of the modes---a distinctive property of the non-degenerate resonance effect. In particular, we find that the radiation from the mode with the larger linewidth preferably takes the total detuning, while the other mode emits photons at its resonance frequency.

\end{abstract}
\maketitle

\section{Introduction}
Parametric down-conversion\cite{guckenheimer} is an important building block of many optical quantum information applications\cite{Gisin:02,Scarani:05}. It describes the process that converts a pump photon of frequency $\Omega$ into a pair of photons with frequencies $\Omega_a$ and $\Omega_b$ which add up to the frequency of the pump photon, $\Omega_a+\Omega_b=\Omega$. This operation can be implemented by nonlinear oscillators with resonance frequencies $\Omega_a$ and $\Omega_b$ that are ac-driven below an instability threshold set by the nonlinear parameters of the oscillator. Parametric down-conversion has been employed to generate entangled photon pairs\cite{Ou:92} as well as squeezed states of light.\cite{Wu:86}

In most optical experiments, only a small fraction of the emitted pairs is detected due to a limited collection efficiency. Therefore, the unconditional emission events appear uncorrelated. However, recent experiments\cite{Astafiev:07,Groot:10,hofheinz:11,Hoi:11,Liu:14,Chen:14} have advanced quantum non-linear optics into the microwave frequency range. In particular, the energy pumped into an electromagnetic environment by a dc-biased Josephson junction is of special interest, as all of the energy associated with a tunneling Cooper pair is confined to transmission lines and electrical oscillators\cite{Drummond:90,Padurariu:12,Leppakangas:13,Armour:13,Gramich:13,Souquet:14,Jin:15,Armour:15,Kubala:15,Leppakangas:16,Wustmann:17}. This constitutes a large technical advantage compared to optical experiments as the concentration of the radiation leads to almost perfect detection efficiencies.

\begin{figure}[tb]
	\centering
	\includegraphics{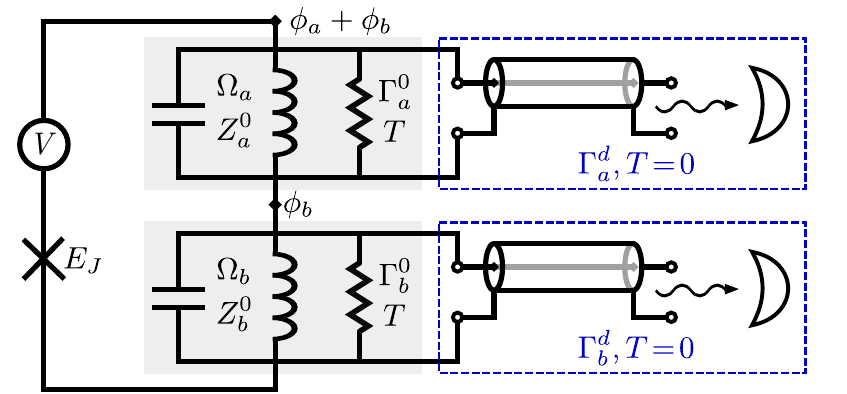}
	\caption{%
		(color online) Setup composed of a Josephson junction with Josephson energy $E_J$ biased by a dc voltage $V$ and coupled to two microwave resonators in series. Each resonator $k \in\{a,b\}$ is characterized by its resonance frequency $\Omega_k$ and an impedance $Z_k^0$. The environment is modeled as resistors at temperature $T$ with a coupling rate $\Gamma_k^0$. Additionally, each resonator is connected to a detector at an effective temperature $T=0$ with a detector rate $\Gamma_k^d$. The setup is characterized by two variables $\phi_a$ and $\phi_b$, which correspond to the superconducting phase difference across each resonator.         
	}\label{fig:setup}
\end{figure}

Recently, accurate measurements of the radiation emitted by a dc-biased Josephson junction in series with two high-quality microwave resonators of different frequencies have demonstrated that such a setup permits the study of correlations.\cite{Westig:17} Motivated by these results, we want to investigate the radiation statistics of the non-degenerate setup illustrated in Fig.~\ref{fig:setup}. The setup is composed of a Josephson junction, dc-biased with voltage $V$, that is in series with two microwave resonators with resonance frequencies $\Omega_a$ and $\Omega_b$. The Josephson frequency can be tuned to the sum of the two resonance frequencies, leading to parametric resonance.

In particular, we want to study the second-order coherences as well as the full counting statistics (FCS) of the resulting radiation. Note that the FCS of the non-degenerate optical parametric oscillator below the instability threshold has been studied previously in Ref.~\onlinecite{Vyas:92} for the specific case of zero detuning between the Josephson frequency and the sum of the two resonance frequencies as well as zero asymmetry in the linewidth of the modes. We revisit the FCS in the limit of the long measurement times employing a different framework based on the Keldysh path integral formalism previously applied in the context of the degenerate parametric oscillator\cite{Padurariu:12}. We extend the earlier results to the case of an arbitrary mismatch between driving frequency and resonance frequencies as well as an arbitrary asymmetry in the linewidth of the modes. We discuss the qualitative differences between the degenerate and the non-degenerate case caused by the interplay of asymmetry and detuning. In particular, we find that the radiation from the mode with the larger linewidth preferably takes the total detuning, while the other mode emits photons at its resonance frequency. Furthermore, we analyze the second-order coherences of the radiation and investigate how thermal effects impact the radiation statistics. 

The paper is organized as follows. We describe the setup in Sec.~\ref{sec:setup}. In Sec.~\ref{sec:keld}, we discuss its Keldysh action. Moreover, we introduce the counting fields which will later provide access to the radiation statistics. In Sec.~\ref{sec:coherence}, we analyze the second-order coherences and discuss the frequency spectrum of the radiation as well as the Fano factors. We include thermal effects in our discussion and calculate the driving strength for which the Cauchy-Schwarz inequality is most strongly violated. We calculate the FCS in Sec.~\ref{sec:fcs} and discuss the FCS in two different limits in Sec.~\ref{sec:limits}. In Sec.~\ref{sec:bursts}, we discuss the interpretation in terms of bursts and calculate the probability of large deviations from the equilibrium. We conclude in Sec.~\ref{sec:conclusion}.

\section{Setup}\label{sec:setup}
The studied setup is composed of a Josephson junction, biased by a dc-voltage source, that is in series with two microwave resonators. The detailed setup is illustrated in Fig.~\ref{fig:setup}. Each resonator $k \in\{a,b\}$ is characterized by its resonance frequency $\Omega_k$ and an impedance $Z_k^0$ at low frequency. The environment is modeled as resistors at temperature $T$ with a coupling rate $\Gamma_k^0$. Additionally, each resonator is connected to a detector. The detectors are assumed to be well isolated such that they effectively operate at $T=0$. They count the radiation emitted by the resonators at a detector rate $\Gamma_k^d$. The photons dissipated by the resistors are not detected causing a finite detection efficiency $f_k$. In total, the emission of photons in each resonator occurs on the frequency scale $\Gamma_k=\Gamma_k^d+\Gamma_k^0$. Since we are primarily interested in correlated events in both resonators, we will normalize time and frequency to the natural scale $\Gamma=(\Gamma_a+\Gamma_b)/2$. 

Close to resonance, the impedance $Z_k$ of resonator $k$ is given by
\begin{equation}
Z_k(\omega)=\frac{Z_k^0\Omega_k}{\gamma_k-i(\omega-\Omega_k)},
\end{equation}
valid for $|\omega-\Omega_k|\ll\Omega_k$. Here, we introduced the abbreviation $\gamma_k=\Gamma_k/(\Gamma_a+\Gamma_b)$.

We are interested in the non-degenerate case, where both resonators have different resonance frequencies, $\Omega_a\neq \Omega_b$. The superconducting phase difference across the junction depends on the superconducting phases $\phi_k$ across each resonator as well as the applied dc voltage $V$. For the following calculations, we require that both resonators have a high quality factor $\Omega_k\gg\gamma_k$. Additionally, we will assume that the impedance far from resonance $Z_k^0$ is small at the quantum scale, $G_QZ_k^0\ll1$, with $G_Q=e^2/\pi\hbar$ the conductance quantum. Then, the contribution to the phase fluctuations from low frequencies $(\delta\phi_k)^2\simeq G_QZ_k^0\ll1$ can be safely neglected. In this context, we also want to neglect the broadening of the Josephson emission line by low-frequency phase noise. This broadening can be estimated as $k_B TG_Q(Z_a^0+Z_b^0)$.\cite{likharev,Arndt:18} The broadening can be neglected, if it is much smaller than the width of both resonators $\hbar\Gamma\gamma_k\gg k_B T G_Q(Z_a^0+Z_b^0)$, which restricts the allowed temperature range of our calculation.

\section{Keldysh action}\label{sec:keld}

We describe the quantum mechanical system using the Keldysh path integral formalism. The partition function $\mathcal{Z}$ is given by the path integral over the Keldysh contour weighted by the factor $e^{iS}$, where $S$ corresponds to the quantum action
\begin{equation}
\mathcal{Z}=\int \mathcal{D}[\phi_a^+]\mathcal{D}[\phi_a^-]\mathcal{D}[\phi_b^+]\mathcal{D}[\phi_b^-]\,e^{iS[\phi_a^+,\phi_a^-,\phi_b^+,\phi_b^-]}.
\end{equation}
Here, $\phi_k^{\pm}(t)$ refers to the superconducting phase difference across the resonators along the forward (backward) part of the Keldysh contour. We divide the action into the contribution from the junction and the resonators, $S=S_J+S_a+S_b$. The action of the Josephson junction is given by an integral of the phase-dependent Josephson energy $U(\phi)=-E_J\cos\phi$ and reads
\begin{align}\label{eq:actionJ}
S_J=\frac{E_J}{\hbar\Gamma}\int \!dt\Big\lbrace\!\cos &\!\big[\phi_V(t)+\phi_a^+(t)+\phi_b^+(t)\big]\nonumber \\
&-\cos\! \big[\phi_V(t)+\phi_a^-(t)+\phi_b^-(t)\big]\Big\rbrace,
\end{align}
with $\phi_V(t)=2eVt/\hbar\Gamma$ the phase difference across the voltage source.

In frequency space, with $\phi_k(\omega)=\int dt \,e^{i\omega t}\phi_k(t)$, the action of resonator $k$ with impedance $Z_k(\omega)$ is given by
\begin{align}
S_k&=\frac{i}{8\pi G_Q}\int \frac{d\omega}{2\pi}\sum_{\alpha,\beta=\pm}\phi_k^{\alpha}(\omega)^{\ast}M_k^{\alpha\beta}(\omega)\,\phi_k^{\beta}(\omega),\\
M_k(\omega)\!&=\!\omega\,\Bigg\lbrace\!\mathrm{Re}\! \left[Z_k(\omega)^{-1}\right]\!\!\begin{bmatrix}
0 \!& 1\\
-1 \!& 0
\end{bmatrix}\!\!+\!i\mathrm{Im} \!\left[Z_k(\omega)^{-1}\right]\!\!\begin{bmatrix}
1 \!& 0\\
0 \!& -1
\end{bmatrix}\nonumber\\
&\quad+\![2n_T(\omega)+1]
\mathrm{Re} \! \left[Z_k(\omega)^{-1}\right] (1-\alpha_k) \begin{bmatrix}
1 \!& -1\\
-1 \!& 1
\end{bmatrix}\nonumber\\
&\quad+\![2n_{0}(\omega)+1]\mathrm{Re}\! \left[Z_k(\omega)^{-1}\right] \alpha_k \!\begin{bmatrix}
1 \!& -1\\
-1 \!& 1
\end{bmatrix}\!\Bigg\rbrace ,
\end{align}
with $n_T(\omega)=[\exp (\hbar\omega \Gamma/k_B T)-1]^{-1}$ the Bose-Einstein occupation at temperature $T$. Additionally, we introduced the factor $\alpha_k=\Gamma_k^d/\Gamma_k$, which originates from the current division between the detector and the resistor. 

To observe the parametric resonance effect, we tune the Josephson frequency $\Omega_J$ close to the sum of the resonator frequencies by setting the dc-bias voltage to $V=\hbar\Gamma\Omega_J/2e$, with $\Omega_J=\Omega_a+\Omega_b+\Delta$. The detuning $\Delta$ is assumed to be small, $\Delta\ll\Omega_k$. Under this assumption\cite{commut}, we can perform a rotating-wave approximation by introducing the slow, complex variables $\psi_k$
\begin{equation}\label{eq:RWA}
 \phi_k^{\pm}(t)=\mathrm{Re}[e^{-i(\Omega_k+\Delta/2) t}\psi_k^{\pm}(t)].
\end{equation}
Note, that the choice of how we distribute the detuning $\Delta$ between both rotations has no physical consequence and corresponds simply to a total frequency shift in our calculations. 

By inserting Eq.~\eqref{eq:RWA} in Eq.~\eqref{eq:actionJ} and neglecting all fast-oscillating terms, we obtain a new action for the slow variables $\psi_k(t)$
\begin{align}
&S_J\!=\!-\frac{E_J}{2\hbar\Gamma}\!\!\int \!dt\, \{ s_J[\psi_a^+(t),\psi_b^+(t)]\!-\!s_J[\psi_a^-(t),\psi_b^-(t)]\},\nonumber\\
&s_J(\psi_a,\psi_b)=\frac{J_1(|\psi_a|)J_1(|\psi_b|)}{|\psi_a||\psi_b|}(\psi_a\psi_b+\psi_a^{\ast}\psi_b^{\ast}),\label{eq:sjfull}
\end{align}
with the Bessel functions $J_{m}(x)$. Similarly, the action of resonator $k$ can be expressed as 
\begin{align}
S_k=\frac{i}{16\pi G_Q}\int &\frac{d\nu}{2\pi}\sum_{\alpha,\beta=\pm}\psi_k^{\alpha}(\nu)^{\ast}\overbar{M}_k^{\alpha\beta}(\nu)\,\psi_k^{\beta}(\nu),\\
Z_k^0\overbar{M}_k(\nu)=-i&\left(\nu+\frac{\Delta}{2}\right) \begin{bmatrix}
1 & 0\\
0 & -1
\end{bmatrix}\nonumber\\
&\qquad+\gamma_k
 \begin{bmatrix}
2n_k+1 & -2n_k\\
-2(n_k+1) & 2n_k+1 
\end{bmatrix} ,\label{eq:envrwa}
\end{align}
with $n_k=n_T(\Omega_k)(1-\alpha_k)$ the average number of photons in the mode in thermal equilibrium.

At $T=0$, the 'classical' equations of motion for the slow, complex variables $\psi_k$ correspond to the saddle-point solution of the total action $S$ for $n_a=n_b=0$ and read
\begin{align}\label{eq:classical}
\frac{d\psi_{k}}{dt}=&\left(\! i\frac{\Delta}{2}
-\gamma_{k}\!\right)\!\psi_{k}\!-\!4i\epsilon\gamma_{k}\sigma_k\Bigg[\psi_{\bar{k}}^{\ast}\frac{J_1(|\psi_a|)J_1(|\psi_b|)}{|\psi_a||\psi_b|}
\nonumber\\
&\,\,-\!\psi_{k}\frac{J_1(|\psi_{\bar{k}}|)J_2(|\psi_{k}|)}{2|\psi_{\bar{k}}||\psi_{k}|^2}(\psi_a\psi_b\!+\!\psi_a^{\ast}\psi_b^{\ast})\Bigg].
\end{align}
Here, $\bar{k}$ corresponds to the other resonator. Additionally, we introduced the dimensionless driving strength $\epsilon=2\pi E_J G_Q \sqrt{Z_a^0Z_b^0}/\hbar\Gamma \sqrt{\gamma_a\gamma_b}$ as well as the abbreviation $\sigma_k=(Z_{k}^0\gamma_{\bar{k}}/Z_{\bar{k}}^0\gamma_k)^{1/2}$. By linearizing Eq.~\eqref{eq:classical}, we find that the classical equations have only nontrivial, nonzero solutions $\psi_{k}\neq 0$ if the Josephson energy exceeds the threshold $\epsilon\geq \epsilon_{\ast}=\sqrt{1+\Delta^2}$.\cite{Wustmann:17} Above this threshold, self-sustained parametric oscillations emerge, leading to coherent radiation from the two oscillators. The two radiation frequencies are shifted from the resonance frequencies due to the detuning $\Delta$. Just above the threshold, the radiation from resonator $k$ is detuned by $\gamma_{k} \Delta$. Therefore, the mode with the larger linewidth is influenced by the detuning more strongly. We will come back to this property later, when we discuss the frequency distribution of photons below the threshold. 

To better understand the behavior of the system above the threshold, we want to calculate the equilibrium positions that correspond to the self-sustained oscillations. Using analogous steps to Ref.~\onlinecite{Armour:15}, we introduce amplitude-phase coordinates for both oscillators, $\psi_{a}=Ae^{i\xi_a}$ and $\psi_{b}=Be^{i\xi_b}$, and expand Eq.~\eqref{eq:classical} to leading order in $\epsilon-\epsilon_*$. In addition to the trivial solution $A=B=0$, we obtain the fixed point positions
\begin{align}\label{eq:A0}
&A=\frac{\sqrt{8\epsilon_*}\sqrt{\epsilon-\epsilon_*}\sigma_a}{\sqrt{\epsilon_*^2(\sigma_a^2+1)\!+\!2\Delta^2(\gamma_b+\sigma_a^2\gamma_a)}},\quad B=\frac{A}{\sigma_a},\\
&e^{i(\xi_a+\xi_b)}=\frac{\Delta-i}{\epsilon_*}.
\end{align}
The fixed point value for the relative phase $\xi_a-\xi_b$ is arbitrary, demonstrating that the phase-locking between the Josephson oscillations and the resonator modes only locks the total phase. This property is qualitatively different to the degenerate parametric oscillator\cite{Wustmann:13}, where the degeneracy of the phase is only two-fold and the phase can take two values differing by $\pi$.

While classical, coherent radiation is no longer possible below the threshold, quantum fluctuations enable the emission of photon pairs. Due to the low impedance environment, phase fluctuations typically remain small, $(\delta\psi_k)^2\simeq Z_k^0G_Q\ll1$. Upon approaching the threshold, the fluctuations increase. However, for sufficiently small impedances, the crossover region where the fluctuations become of the order of $1$ remains narrow and can be estimated as $|\epsilon_*-\epsilon|\simeq Z_{k}^0G_Q\gamma_{\bar{k}}\epsilon_*\ll\epsilon_*$. Therefore, we can expand the Josephson action in Eq.~\eqref{eq:sjfull} up to leading order in $\psi_k$ 
\begin{align}
S_J=-\frac{E_J}{8\hbar\Gamma}\!\!\int \!\!dt\,[\psi_a^{+}(t)\psi_b^+(t)-\psi_a^{-}(t)\psi_b^-(t)+\mathrm{c.c.}],\end{align}
valid up to a narrow interval below the threshold.

Next, we want to introduce the counting fields $\chi_k(t)$ which will later provide access to the radiation statistics. The counting fields count only the photons emitted at the detectors, which are effectively at zero temperature. A common choice is a piecewise-constant counting field, with $\chi_k(t)=\chi_k$ during the time interval when the respective detector counts the emitted photons and $\chi_k(t)=0$ at all other times. We can include the counting fields in our calculation by modifying the second term in Eq.~\eqref{eq:envrwa} into\cite{chi}
\begin{align}
\gamma_k
 \begin{bmatrix}
2n_k+1 & -2n_k\\
-2(n_k+1)[1+f_k(e^{i\chi_k}-1)] & 2n_k+1
\end{bmatrix}.\label{eq:chimod}
 \end{align}
Here, we have introduced the counting efficiency $f_k=\alpha_k/(n_k+1)$ which is generally finite due to additional dissipation at the resistors at temperature $T$. With the modification in Eq.~\eqref{eq:chimod}, $\mathcal{Z}(\chi_k)$ represents the characteristic function of the probability distribution of detecting $N_k$ photons within the chosen time interval
\begin{align}
P(N_k)=\int \frac{d\chi_k}{2\pi}\mathcal{Z}(\chi_k)e^{-i\chi_k N_k}.\label{eq:probability}
\end{align}
The total action of the linearized system can be written in compact form by introducing the vector $\Psi(\nu)=[\psi_a^+(\nu),\psi_a^-(\nu),\psi_b^{+}(-\nu)^{\ast},\psi_b^{-}(-\nu)^{\ast}]^T$. In this basis, the action reads
\begin{align}
S=\frac{i}{16\pi G_Q}\!\int &\frac{d\nu}{2\pi}\sum_{\alpha,\beta}\Psi^{\alpha}(\nu)^{\ast}A^{\alpha\beta}(\nu,\chi_a,\chi_b)\,\Psi^{\beta}(\nu),\label{eq:amatrixdef}\\
A(\nu,\chi_a,\chi_b)=& \begin{bmatrix}
\overbar M_a(\nu,\chi_a) & J\\
J & \overbar M_b^T(-\nu,\chi_b)
\end{bmatrix}.\label{eq:amatrix}
\end{align}
Here, $\overbar M_k(\nu,\chi_k)$ is given by Eq.~\eqref{eq:envrwa} with the modification in Eq.~\eqref{eq:chimod} and 
\begin{align}
J=\frac{i2\pi E_J G_Q}{\hbar\Gamma}\begin{bmatrix}
1 & 0\\
0 & -1\end{bmatrix}.
\end{align}

\section{Second-order coherences}\label{sec:coherence}
In this section, we want to calculate the second-order coherences 
\begin{align}
g^{(2)}_{kl}(\tau)=\frac{\langle:\!I_{l}(\tau)I_k(0)\!:\rangle}{\langle I_k\rangle\langle I_{l}\rangle},\quad k,l\in\lbrace a,b\rbrace
\end{align}
of the radiation. Here, we introduced the photon current $I_k(\tau)$ in the detector from resonator $k$. In our calculation, the Keldysh path integral formalism guarantees the contour-ordering of the operators due to the inherent time-ordering along the Keldysh contour.\cite{beenakker:01,hassler:15} 

Let us choose the time-dependent, piecewise constant counting fields as follows. During the time interval $(0,\Delta t)$, we count the photons emitted by resonator $k$ and therefore set $\chi_k(t)=\chi_k$. Within the time interval $(\tau,\tau+\Delta t)$, we count the photons emitted by resonator $l$ and set $\chi_l(t)=\chi_{l}$. At all other times, the counting fields are set to zero. We assume that the measurement time $\Delta t$ is small, such that the average number of photons emitted is small, $ \Delta t \ll1$. Additionally, we assume that the time distance between the measurement intervals is large, $\tau\gg \Delta t$. 

We can calculate the number of photons $\Delta\langle N_{k}\rangle$ detected in the small time interval $\Delta t$ with the characteristic function via $\Delta\langle N_{k}\rangle=d \mathcal{Z}/d (i\chi_k)|_{\chi_k=0}$. We obtain
\begin{align}\label{eq:Ncorr}
\bar{I}_k=\!\frac{ f_k \gamma_k(n_k\!+\!1)}{8\pi G_Q \!Z_k^0}\langle\psi_k^{-\ast}\psi_k^{+}\!\rangle ,
\end{align}
with $\bar{I}_k=\Delta\langle N_{k}\rangle/\Delta t$.
Since the action is Gaussian, the calculation of the correlators is straightforward. To avoid cluttering the main article, the detailed derivation can be found in App.~\ref{app:correlator}. We insert the correlator in Eq.~\eqref{eq:Ncorr} and obtain
\begin{align}
\bar{I}_k=2 f_k\gamma_{k}\frac{1+n_{k}}{\epsilon_*^2-\epsilon^2}\left\lbrace n_{k}\epsilon_*^2
\!+\!\epsilon^2\!\left[(1\!+\!n_{\bar{k}})\gamma_{\bar{k}}\!-\!n_{k}\gamma_{k}\right]\right\rbrace\!.\label{eq:intensity}
\end{align}
It is instructive to express this intensity as an integral of the frequency spectrum of the radiation $\rho(\nu)$. At $T=0$, we obtain
\begin{align}
\bar{I}_k=&\  f_k\int\frac{d\nu}{2\pi}\,\rho(\nu),\qquad \rho(\nu)=\frac{(1-\gamma^2)^2\epsilon^2}{4p(\nu)},\label{eq:rhoomega}
\\
p(\nu)= & \ \nu ^4 +\frac{1+r}{2} \nu
  ^2 - \nu 
  \gamma \Delta + \frac{(1-r)^2+4 \gamma ^2 \Delta
 ^2}{16}.\label{eq:pomega}\end{align}
Here, we have introduced the parameter
\begin{align}
 r &= (1-\gamma^2) \epsilon^2 + \gamma ^2 -\Delta ^2, 
\end{align}
where $r$ ranges from $\gamma^2-\Delta^2$ at zero drive to $1-\gamma^2\Delta^2$ at the threshold. Additionally, we have introduced the asymmetry 
\begin{equation}
 \gamma = \gamma_a-\gamma_b,\qquad |\gamma|
 < 1.
\end{equation}
\begin{figure}[tb]
	\centering
	\includegraphics{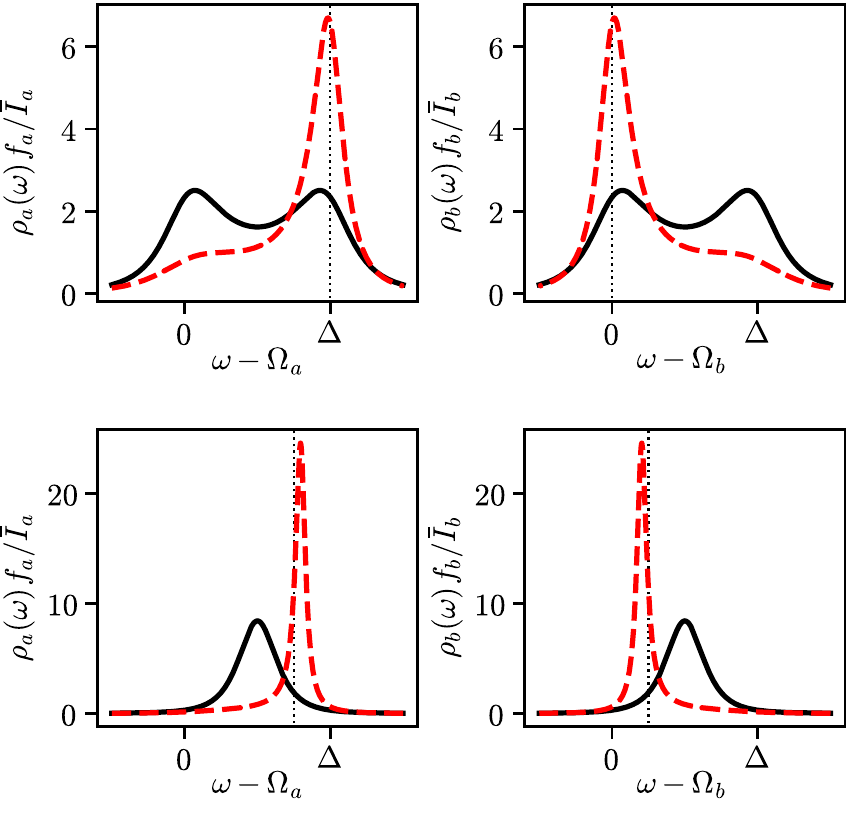}
	\caption{%
		(color online) Frequency spectrum of the radiation $\rho_k(\omega)$ from resonator $a$ (left) and $b$ (right) for $\Delta=2$ and $\epsilon/\epsilon_*=0.1$ (top), $0.9$ (bottom). Here, $\rho_k(\omega)$ corresponds to $\rho(\nu)$ in Eq.~\eqref{eq:rhoomega} with $\nu=\pm(\omega-\Omega_k-\Delta/2)$. The figures for resonator $b$ can be obtained by mirroring the figures for resonator $a$ at $\Delta/2$. The black, solid lines correspond to zero asymmetry ($\gamma=0$). The red, dashed lines correspond to $\gamma=0.5$, where the linewidth of resonator $a$ is three times as large as the linewidth of resonator $b$. At low driving strength (top), the resonators are most likely to emit photons either at their resonance frequency $\Omega_k$ or at the detuned frequency $\Omega_k+\Delta$. Which resonator preferably takes the total detuning is determined by the asymmetry. The dotted lines correspond to a radiation at $\Omega_a+\Delta$ and $\Omega_b$. Close to the threshold (bottom), the detuning is split between both resonators according to the linewidth, such that the radiation from resonator $k$ is detuned by $\gamma_{k} \Delta$ as indicated by the dotted lines. 
	}\label{fig:spectral_density}
\end{figure}
Figure~\ref{fig:spectral_density} displays the emission spectrum as a function of frequency. At low driving strength, $\epsilon\ll\epsilon_*$, the resonators are most likely to emit photons either at their resonance frequency $\Omega_k$ or at the detuned frequency $\Omega_k+\Delta$. The detuning is not split between the resonators. Which resonator preferably takes the total detuning is determined by asymmetry. At zero asymmetry, both resonators are equally likely to emit photons at the detuned frequency. However, in the presence of asymmetry, the radiation from the mode with the larger linewidth is more likely to contain the total detuning. This is a distinct feature of the non-degenerate oscillator. 

Close to the threshold, $\epsilon\approx\epsilon_*$, the detuning is split between both resonators such that the radiation from resonator $k$ is detuned by $\gamma_{k} \Delta$. This behavior coincides with the classical results discussed in the previous section. 

 \begin{figure}[tb]
	\centering
	\includegraphics{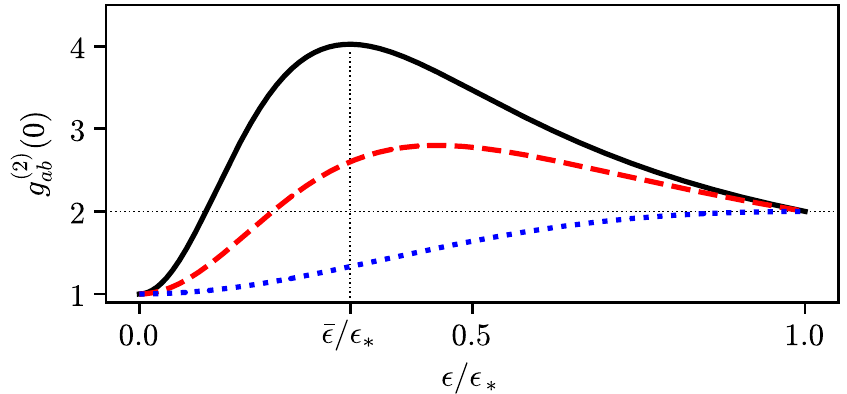}
	\caption{%
		(color online) Second-order coherence $g_{ab}^{(2)}(0)$ as a function of driving strength $\epsilon$ for $\gamma=0$. The lines correspond to varying photon occupations in thermal equilibrium with $n_a\!=\!n_b\!=0.05$ (black, solid), $n_a\!=\!n_b\!=0.1$ (red, dashed), and $n_a\!=\!n_b\!=\!0.5$ (blue, dotted). The horizontal line indicates the value of $g_{aa}^{(2)}(0)=g_{bb}^{(2)}(0)=2$. For $g_{ab}^{(2)}(0)>2$, the Cauchy-Schwarz inequality is violated. Maximum violation is achieved for $\epsilon=\bar{\epsilon}$. The dependence of this maximum value on asymmetry is illustrated in Fig.~\ref{fig:g2abmax}.
	}\label{fig:g2ab0}
\end{figure}
Analogously to Eq.~\eqref{eq:Ncorr}, we can calculate the second-order correlator
\begin{align}
\langle:\!I_{l}(\tau)I_k(0)\!:\rangle&=\frac{1}{(\Delta t)^2}\frac{d^2 \mathcal{Z} }{d (i\chi_k)d (i\chi_{l})}\!\Big|_{\chi_k,\chi_{l}=0}\nonumber\\
&\propto\langle\psi_k^- (0)^{\ast}\psi_l^{-}(\tau)^{\ast}\psi_{l}^{+}(\tau)\psi_k^{+}(0)\rangle .
\end{align}
By applying Wick's theorem, we can calculate the correlator using the results from App.~\ref{app:correlator}. Close to threshold, $\epsilon\approx\epsilon_*$, the coherences $g_{aa}$, $g_{bb}$, and $g_{ab}$ all demonstrate the same asymptotic behavior. We obtain $g_{kl}^{(2)}(\tau)=1+\exp (-|\tau|/\tau_{\epsilon})$ with
\begin{align}
\tau_{\epsilon}=\frac{1+\gamma^2\Delta^2}{\epsilon_*(\epsilon_*-\epsilon)(1-\gamma^2)}.\label{eq:g2time}
\end{align}
The expression suggests the emergence of a new long time scale $\tau_{\epsilon}$ in the vicinity of the threshold which diverges upon approaching the threshold. This behavior is similar to that of the degenerate oscillator \cite{Padurariu:12}. However, for the non-degenerate oscillator asymmetry can have a strong influence on the time scale.

\begin{figure}[tb]
	\centering
	\includegraphics{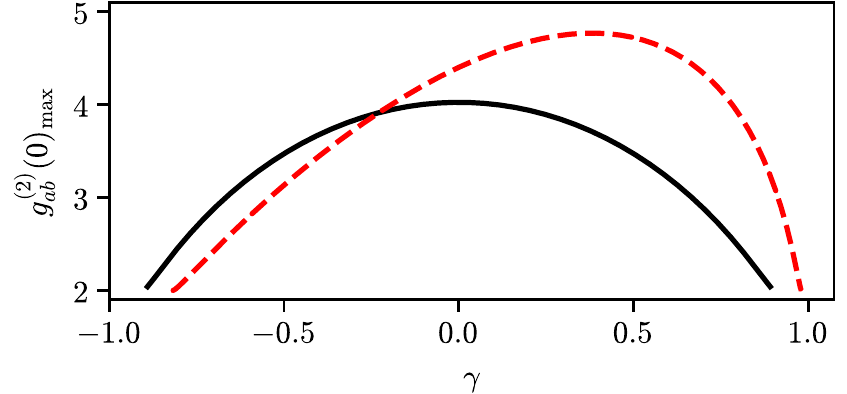}
	\caption{%
		(color online) Value of $g_{ab}^{(2)}(0)$ at the optimal drive $\epsilon=\bar{\epsilon}$ as a function of asymmetry $\gamma$. The black, solid lines corresponds to $n_a=n_b=0.05$. The red, dashed line corresponds to $n_a=0.01$ and $n_b=0.1$. If the occupation numbers of both resonators in thermal equilibrium coincide, the maximum coherence is obtained for zero asymmetry and the graph is symmetric around $\gamma=0$. However, asymmetry increases the coherence, if the occupation numbers in thermal equilibrium differ.
	}\label{fig:g2abmax}
\end{figure}

In addition to the exponential behavior, the second-order coherences for $\tau=0$ are also of interest. Figure~\ref{fig:g2ab0} displays the second-order coherences for $\tau=0$ as a function of driving strength. Analytically, we obtain
\begin{align}
&g_{aa}^{(2)}(0)=g_{bb}^{(2)}(0)=2,\\
&g_{ab}^{(2)}(0)\!\overset{\text{$\epsilon\!\ll\!\bar{\epsilon}$}}{=}1+\frac{\epsilon^2}{\epsilon_*^2}\frac{(1+n_a+n_b)^2\gamma_a\gamma_b}{n_a n_b},\\
&g_{ab}^{(2)}(0)\!\overset{\text{$\epsilon\!\approx\!\epsilon_*$}}{=}2+\frac{2(\epsilon_*-\epsilon)}{\epsilon_*(1\!+\!n_a\!+\!n_b)\!}\!\left(\!\!1\!-\!n_a\frac{\gamma_a}{\gamma_b}\!-\!n_b\frac{\gamma_b}{\gamma_a}\!\right)\!.\label{eq:gab0}
\end{align}
Here, $\bar{\epsilon}$ is given by
\begin{align}
\bar{\epsilon}=\epsilon_*\!\left\lbrace\!\frac{n_a n_b}{[(1+n_a)\gamma_a-n_b\gamma_b][(1+n_b)\gamma_b-n_a\gamma_a]}\!\right\rbrace^{\!1/4}
\end{align}
and corresponds to the drive at which $g_{ab}^{(2)}(0)$ reaches its maximum value of
\begin{align}
g_{ab}^{(2)}(0)\!\overset{\text{$\epsilon\!=\!\bar{\epsilon}$}}{=}1+\frac{\gamma_a\gamma_b}{(\sqrt{\gamma_a n_a}+\sqrt{\gamma_b n_b}\,)^2},\label{eq:gab0max}
\end{align}
valid for small temperature such that $n_a,n_b\ll1$. Figure~\ref{fig:g2abmax} displays the maximum value of $g_{ab}^{(2)}(0)$ as a function of asymmetry. If the occupation numbers of both resonators in thermal equilibrium coincide, asymmetry decreases coherence. However, since the resonance frequencies of the resonators differ, the occupation numbers in thermal equilibrium differ, too. In this case, asymmetry can increase the coherence. To achieve maximal coherence, the resonator with the larger resonance frequency and, thus, the lower photon occupation in thermal equilibrium should feature the larger linewidth.
Then, the Cauchy-Schwarz inequality
\begin{align}
\sqrt{g_{aa}^{(2)}(0)g_{bb}^{(2)}(0)}\geq g_{ab}^{(2)}(0)
\end{align}
is most strongly violated. The violation of the inequality is linked to nonclassical correlations between the radiation from the two resonators and demonstrates the quantum character of the radiation. At $T=0$, the Cauchy-Schwarz inequality is violated everywhere below the threshold. Due to thermal processes at finite temperature, the inequality is no longer violated for all driving strength. Instead, the driving strength needs to exceed $\epsilon\geq\bar{\epsilon}^2/\epsilon_*$ to generate nonclassical radiation. Furthermore, Eq.~\eqref{eq:gab0} indicated that for $n_a\gamma_a/\gamma_b+n_b\gamma_b/\gamma_a\geq1$, the Cauchy-Schwarz inequality is always fulfilled for any driving strength.

The second-order coherences can also be used to calculate the Fano factor $F_{kk}$. The Fano factor is connected to the second-order coherence via\cite{hassler:15}
\begin{align}
F_{kk}=1+\bar{I}_{k}\int d\tau \left[g_{kk}^{(2)}(\tau)-1\right]
\end{align}
and corresponds to the number of correlated photons from either resonator. We obtain
\begin{align}
F_{kk}\!\overset{\text{$\epsilon\!\ll\!\bar{\epsilon}$}}{=}\,&1\!+\!2 f_{k}n_k\!+\!\frac{2\epsilon^2f_k\gamma_{\bar{k}}}{\epsilon_*^4}\big[n_k\epsilon_*^2\nonumber\\
&\qquad \ +\!\big(1\!+\!2n_k\!+\!n_{\bar{k}}\big)\big(4\gamma_k\!+\!\epsilon_*^2\gamma_{\bar{k}}\!-\!\epsilon_*^2\gamma_k\big)\big],\\
F_{kk}\!\overset{\text{$\epsilon\!\approx\!\epsilon_*$}}{=}\,&\frac{f_k(1\!+\!\gamma^2\Delta^2)}{2(\epsilon_*\!-\!\epsilon)^2}\left(1\!+\!2n_k\!+\!n_{\bar{k}}\right),
\end{align}
valid for low temperature, such that $n_a,n_b\ll1$. In the limit $\epsilon\rightarrow 0 $, we obtain the Fano factor expected from thermal radiation\cite{Brange:19}. Upon approaching the threshold, the Fano factor diverges quadratically. Furthermore, we know from Eq.~\eqref{eq:Ncorr} that the number of photons present in either resonator diverges only linearly upon approaching the threshold. Thus, the number of correlated photons exceeds by far the number of photons present in either resonator. This behavior is connected to the long times scale $\tau_{\epsilon}$ in the vicinity of the threshold, see also Sec.~\ref{sec:bursts} and Ref.~\onlinecite{Padurariu:12}.

\begin{figure}[tb]
	\centering
	\includegraphics{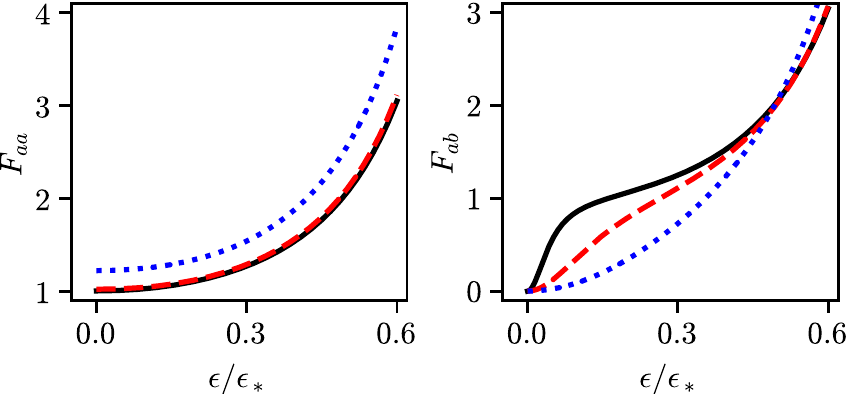}
	\caption{%
		(color online) Fano factors $F_{aa}$ (left) and $F_{ab}$ (right) as a function of driving strength $\epsilon$ for $\Delta=0$, $\gamma=0$, and $f_a=f_b=1$. The lines correspond to varying photon occupations in thermal equilibrium with $n_a\!=\!n_b\!=0.001$ (black, solid), $0.01$ (red, dashed), $0.1$ (blue, dotted). We have checked that a finite $\gamma$ or $\Delta$ does not change the plot qualitatively. Note that thermal effects have a strong impact on $F_{ab}$ at small drive.
	}\label{fig:Faaab}
\end{figure}

Additionally, we can calculate the Fano factor $F_{ab}$, given by\cite{fano}
\begin{align}
F_{ab}=\sqrt{\bar{I}_a\bar{I}_b}\int d\tau \left[g_{ab}^{(2)}(\tau)-1\right],
\end{align}
which corresponds to the number of correlated photons across resonators. We obtain
\begin{align}
F_{ab}\!\overset{\text{$\epsilon\!\ll\!\bar{\epsilon}$}}{=}\,&\sqrt{\frac{f_{a}f_{b}\gamma_a\gamma_b}{n_a n_b}}\frac{\epsilon^2}{\epsilon_*^2},\\
F_{ab}\!\overset{\text{$\epsilon\!\approx\!\epsilon_*$}}{=}\,&\frac{\sqrt{f_{a}f_{b}}(1\!+\!\gamma^2\Delta^2)}{2(\epsilon_*\!-\!\epsilon)^2}\left(1\!+\!\frac{3}{2}n_a\!+\!\frac{3}{2}n_b\right),
\end{align}
valid for $n_a,n_b\ll1$. Figure~\ref{fig:Faaab} displays both Fano factors as a function of driving strength. 
Close to threshold, the behavior of $F_{ab}$ is similar to that of $F_{kk}$. In fact, for $f_a=f_b$, we find $F_{ab}=(F_{aa}+F_{bb})/2$. In contrast, the behavior of $F_{ab}$ at small drive, $\epsilon\ll\bar{\epsilon}$, differs from the behavior of $F_{kk}$ and depends strongly on thermal effects. As expected, at zero drive, there is no correlation between resonator $a$ and $b$. However, at small temperatures such that $n_a,n_b\ll1$, even a small drive can strongly increase $F_{ab}$. In the limit $T\rightarrow 0$, the number of correlated photons jumps from $0$ at zero drive to $\sqrt{f_{a}f_{b}}$ at finite drive. Thus, at $T=0$ and finite drive, a photon emitted from resonator $k$ is always correlated with at least one photon emitted from resonator $\bar{k}$. However, the jump is very sensitive to thermal effects and easily washed out as displayed in Fig.~\ref{fig:Faaab}.

\section{Full counting statistics}\label{sec:fcs}

This section focuses on the calculation of the full counting statistics (FCS). Since we already discussed the influence of thermal radiation in the previous section, we will focus on the limit $T=0$. We count the photons emitted from either resonator in the measurement interval $(0,\tau)$. Analogously to the previous section, we set $\chi_k(t)=\chi_k$ during the measurement interval and $\chi_k(t)=0$ at all other times. Additionally, we introduce discrete frequencies spaced equally apart with $2\pi/\tau$. Since the action separates for each discrete frequency, evaluating the path integrals amounts to calculating the determinant of the matrix $A(\nu,\chi_a,\chi_b)$ at each frequency. We turn the resulting product of determinants into the exponent of a sum. In the long time limit, we can transform the sum back to a frequency integral and obtain
\begin{align}
\mathcal{Z}(\chi_a,\chi_b)=e^{\lambda (\chi_a,\chi_b)}.
\end{align}
Here, the cumulant-generating function is given by
\begin{align}
\lambda(\chi_a,\chi_b)&=- \tau\int \frac{d\nu}{2\pi}\ln\left[\frac{\det A(\nu,\chi_a,\chi_b)}{\det A(\nu,\chi_a\!=\!\chi_b\!=\!0)}\right]\nonumber\\
&=- \tau \int\frac{d\nu}{2\pi}
 \ln\left[1 - \rho(\nu)s(\chi_a,\chi_b)\right],\label{eq:lambdaint}
\end{align}
with
\begin{align}
 s(\chi_a,\chi_b) =\ & (e^{i\chi_a} -1)f_a(1-f_b)+(e^{i\chi_b} -1)f_b(1-f_a) \nonumber\\
 &+ (e^{i(\chi_a+\chi_b)} -1)f_af_b\label{eq:schi}
 \end{align}
and the frequency spectrum of the radiation $\rho(\nu)$ defined in Eq.~\eqref{eq:rhoomega}. The structure of Eq.~\eqref{eq:schi} suggests that photons are always emitted in pairs. This is consistent with the results from the previous section. A photon emitted from resonator $k$ is always correlated with at least one photon emitted from resonator $\bar{k}$. However, the detection of a single, seemingly uncorrelated photon is possible due to a finite detection efficiency. 

Following the steps presented in App.~\ref{app:integral}, the integral in Eq.~\eqref{eq:lambdaint} can be evaluated to obtain
\begin{equation}\label{eq:lambdam}
 \lambda(\chi_a,\chi_b) =  \tau \Big(1-\sqrt{m} \Big),
\end{equation}
where $m$ is the solution of
\begin{equation}\label{eq:cubic}
 (m-1) \Bigl[ m (r-m) + \gamma^2 \Delta^2 \Bigr] = (1-\gamma^2)^2 \epsilon^2s m 
\end{equation}
that coincides with $1$ for $s=s(\chi_a,\chi_b)=0$. This is one of the main results of this paper. 
The full analytic expression of $m$ is given in App.~\ref{app:fullm}. 

We can use Eq.~\eqref{eq:cubic} to obtain the derivatives
\begin{align}
 \frac{dm}{ds}\Big|_{s=0}\!\! &= - \frac{(1-\gamma^2) \epsilon^2}{\epsilon_*^2-\epsilon ^2}\label{eq:firstder}\\
 \frac{d^2m}{ds^2}\Big|_{s=0} \!\!&= -
 \frac{ 2(1+\gamma ^2 \Delta ^2) (1-\gamma^2) \epsilon^4}
 { (\epsilon_* ^2-\epsilon ^2)^3}.\end{align}
These derivatives are useful to obtain explicit expressions for the various cumulants. The average intensity $\bar{I}_k =\langle N_k\rangle/\tau$ is given by
\begin{equation}
 \frac{\langle N_k\rangle}{\tau}=\frac{1}{\tau}\frac{d \lambda}{d (i\chi_k)}\Big|_{\chi_a,\chi_b=0}=\frac{ f_k \epsilon^2 (1-\gamma^2)}{2(\epsilon_*^2-\epsilon^2)},
\end{equation}
which coincides with our previous results in Eq.~\eqref{eq:intensity} at $T=0$. The intensity noise $S_{k}=\langle\langle N_k^2\rangle\rangle/\tau$ is given by 
\begin{align}
 \frac{\langle\langle N_k^2\rangle\rangle}{\tau}&=\frac{1}{\tau}\frac{d^2 \lambda }{d (i\chi_k)^2}\Big|_{\chi_a,\chi_b=0}= \frac{\langle N_k\rangle}{\tau}\nonumber\\&\qquad\quad\ \ \,+ \frac{ f_k^2\epsilon^4(5-r+3\gamma^2\Delta^2)(1-\gamma^2)}{4(\epsilon_*^2-\epsilon^2)^3}.
\end{align}
Analogously, the cross-resonator noise $S_{ab}=\langle\langle N_a N_b\rangle\rangle/\tau$ is given by 
\begin{align}
 \frac{\langle\langle N_a N_b\rangle\rangle}{\tau}&=\frac{1}{\tau}\frac{d^2 \lambda }{d (i\chi_a)d (i\chi_b)}\Big|_{\chi_a,\chi_b=0}\!\!= \frac{\sqrt{f_af_b\langle N_a\rangle\langle N_b\rangle}}{\tau}\nonumber\\&\,\,\,\,\,\quad+ \frac{ f_a f_b\epsilon^4(5-r+3\gamma^2\Delta^2)(1-\gamma^2)}{4(\epsilon_*^2-\epsilon^2)^3}.
\end{align}
The Fano factors previously discussed in Sec.~\ref{sec:coherence} can also be obtained directly from these results with $F_{kk}=S_k/\bar{I}_k$ and $F_{ab}=S_{ab}/(\bar{I}_a\bar{I}_b)^{1/2}$. 

\section{Limits}\label{sec:limits}
In this section, we briefly discuss the FCS in two different limits.
\subsection{No detuning or no asymmetry}
In the cases where either $\Delta =0 $ (no detuning) or $\gamma=0$ (no asymmetry), Eq.~\eqref{eq:cubic} simplifies and we find the simple explicit expression
\begin{equation}
 m = \frac{1}{2} \Bigl[1 + r + \sqrt{(1-r)^2
 -4 (1-\gamma^2)^2 \epsilon^2 s} \Bigr].
\end{equation}
The resulting expression for the cumulant-generating function coincides with the results for the optical, non-degenerate oscillator in Ref.~\onlinecite{Vyas:92} for $\tau\to\infty$. Additionally, the result is of similar form as the FCS of the degenerate oscillator in Ref.~\onlinecite{Padurariu:12} as well as the optical result in Ref.~\onlinecite{Vyas:89} for $\tau\to\infty$. In this limit, the non-degenerate oscillator displays the same qualitative characteristics as the degenerate oscillator. This indicates that asymmetry can only have a qualitative effect on the statistics of the radiation, if the detuning is finite. We already discovered such an effect in Sec.~\ref{sec:coherence}, when we discussed how the detuning is distributed between both resonators.

\subsection{Close to threshold}
In the vicinity of the threshold, we can obtain a general expression for the derivatives of $m$ using Eq.~\eqref{eq:cubic} 
\begin{equation}
 \frac{d^nm}{ds^n}\Big|_{s=0} \!\!= - \frac{(2n)!}{2 (2n\!-\!1) n!}
 \frac{(1\!-\!\gamma^2)(1\!+\!
  \gamma^2\Delta^2)^{n-1} \epsilon_*^{2n}}{ (\epsilon_*
 ^2-\epsilon ^2)^{2n -1} }.
\end{equation}
The corresponding function $m$ is given by
\begin{align}
 m=1&+ \frac{(1-\gamma^2)}{2(1+\gamma^2 \Delta^2)} \Big(\epsilon^2 - \epsilon_*^2 \nonumber\\
  &\qquad\quad+\sqrt{(\epsilon_*^2-\epsilon^2)^2 -4 (1+\gamma^2 \Delta^2)\epsilon_*^2 s}\Big).
\end{align}
The result can in fact also be obtained directly by expanding Eq.~\eqref{eq:cubic} around $m=1$ with the result
\begin{equation}
 (m-1) \Bigl[ r+2\gamma^2 \Delta^2 -m (1+ \gamma^2 \Delta^2) \Bigr] = (1-\gamma^2)^2 \epsilon_*^2
 s.
\end{equation}
This expansion stems from the fact that, at the threshold, the cumulant-generating function becomes non-analytic for $s=0$ ($m=1$), see App.~\ref{app:analytic}. The non-analytic point of the cumulant-generating function is also of interest in the next section when we discuss the probability of big deviations.

\section{Bursts and large deviations}\label{sec:bursts}

In this section, we will assume a perfect counting efficiency, $f_a=f_b=1$. In this limit, Eq.~\eqref{eq:schi} is reduced to $s(\chi)=e^{i\chi} -1$ with $\chi=\chi_a+\chi_b$. This shows that photons are always detected in pairs with $\bar{I}_a=\bar{I}_b=\bar{I}$. The form of $\lambda (\chi)$ in Eq.~\eqref{eq:lambdaint} allows for an expansion in $e^{i\chi}$. We obtain
\begin{align}
\lambda (\chi)=\sum_{j=1}^\infty\lambda_j\left( e^{ij\chi }-1\right),\label{eq:lambdabursts}
\end{align}
where the coefficients are given by
\begin{align}
\lambda_j=\frac{\tau}{j}\int \frac{d\nu}{2\pi}\left[ \frac{\rho(\nu)}{1+\rho(\nu)}\right]^j.\label{eq:bursts}
\end{align}
In the vicinity of the threshold, we can calculate the asymptotic behavior of the coefficients analytically.
In the limit $j\gg1$, we can approximate Eq.~\eqref{eq:bursts} by a Gaussian integral and obtain
\begin{align}
\lambda_{j\gg 1}\simeq\frac{\tau(1-\gamma^2)\epsilon_*}{4j^{3/2}\sqrt{\pi(1+\gamma^2\Delta^2)}}e^{-j\mu_0},\label{eq:lambdaj}
\end{align}
where $\mu_0$ corresponds to the minimum of $\ln[1+1/\rho(\nu)]$ as a function of frequency $\nu$. In the vicinity of the threshold $\mu_0$ is given by
\begin{align}\label{eq:mu0}
\mu_0=\frac{(\epsilon_*-\epsilon)^2}{1+\gamma^2\Delta^2}=\frac{(1-\gamma^2)^2\epsilon_*^2}{16\bar{I}^2(1+\gamma^2\Delta^2)}.
\end{align}
A more general expression for $\mu_0$ is given in App.~\ref{app:analytic}. 

The form of the cumulant-generating function in Eq.~\eqref{eq:bursts} suggests that photons are emitted in uncorrelated bursts of $j$ photon pairs. Since this behavior has been previously discussed in the context of the degenerate parametric oscillator\cite{Padurariu:12}, we will not go into more details here.

\begin{figure}[tb]
	\centering
	\includegraphics{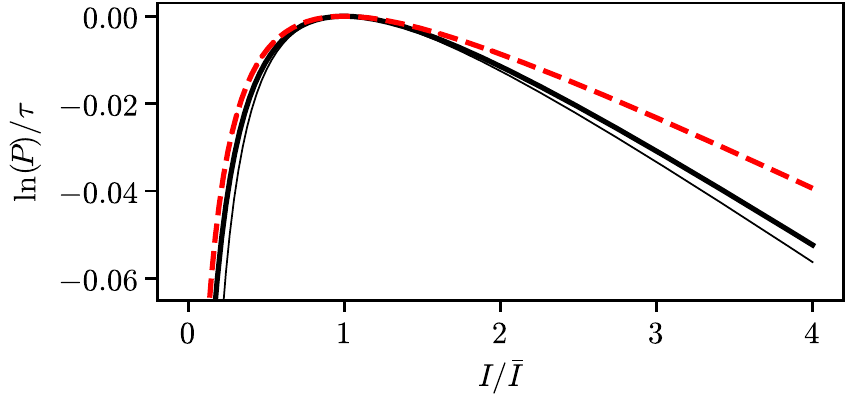}
	\caption{%
		(color online) Probability of large deviations of the photon current $I$ from its average value $\bar{I}$ for $\epsilon=0.9$, $\Delta=0$ and $\gamma=0$ (black, solid), $0.5$ (red, dashed). The thin line indicates the asymptotic approximation in the vicinity of the threshold [Eq.~\eqref{eq:extr}]. Asymmetry increases the probability of large deviations.
	}\label{fig:ExtremeI}
\end{figure}

Instead, we want calculate the probability of large deviations of the photon current $I$ from its average value $\bar{I}$. Using Eq.~\eqref{eq:lambdaj}, we can calculate the corresponding cumulant generating function $\lambda(\chi)$ by approximating the sum in Eq.~\eqref{eq:lambdabursts} by an integral. Then, we evaluate the probability in Eq.~\eqref{eq:probability} at $N=I\tau$ using the saddle-point approximation. To exponential accuracy, we obtain
\begin{align}\label{eq:extr}
P(I)&=\int \frac{d\chi}{2\pi}\exp [-i\chi I\tau+\lambda(\chi)]\nonumber\\
&\propto \exp\! \left[-\frac{\tau}{4\tau_{\epsilon}}\left(\frac{I}{\bar{I}}\!+\!\frac{\bar{I}}{I}\!-\!2\right)\right],
\end{align}
with $\tau_{\epsilon}$ previously defined in Eq.~\eqref{eq:g2time}. The form of this expression is identical to the degenerate oscillator case\cite{Padurariu:12} suggesting a universality of both systems that is connected to the diverging time scale $\tau_{\epsilon}$. Figure~\ref{fig:ExtremeI} displays the probability of large deviations at different driving strengths. Asymmetry increases the probability of large deviations.
\section{Conclusion}\label{sec:conclusion}

In view of recent advances in experimental setups that enable the measurement of correlations, we have studied the statistics of the radiation due to non-degenerate Josephson parametric resonance. We discussed the impact of detuning and asymmetry on the frequency spectrum of the radiation. We found that at low driving strength the detuning is not split between both resonators. Instead, the photons from the resonator with the larger linewidth contain the total detuning. With increasing driving strength, the detuning is split between the resonators resembling the classical behavior above the instability threshold.

Next, we investigated how thermal effects and asymmetry impact the second-order coherence as well as the Fano factor of the emitted radiation. In particular, we were able to obtain the driving strength for which the Cauchy-Schwarz inequality is most strongly violated. Furthermore, we found that the cross-coherence can be increased, if the resonator with the larger resonance frequency has the larger linewidth. Both properties add up to an optimal parameter regime for demonstrating non-classical correlations.

In addition, we calculated the full counting statistics of the radiation. We discovered that in the limit of no asymmetry or no detuning the full counting statistics of the non-degenerate oscillator can be mapped back to the degenerate oscillator. Our results are valid below the threshold where phase fluctuations are small. In the future, it would be interesting to investigate the radiation statistics in the narrow crossover region at the instability threshold where phase fluctuations increase. 

\appendix
\section{Derivation of correlators}\label{app:correlator}
For the calculations presented in the main text, we need the propagators
\begin{equation}
G_{\alpha\beta}(\tau)=\langle\Psi^{\alpha}(\tau)\Psi^{\beta}(0)^{\ast}\rangle=16\pi G_Q(A^{\alpha\beta})^{-1}(\tau),
\end{equation}
where the last step follows from Gaussian integration of Eq.~\eqref{eq:amatrixdef} with $(A^{\alpha\beta})^{-1}(\tau)$ defined via inverse Fourier transform
\begin{equation}
(A^{\alpha\beta})^{-1}(\tau)=\int\frac{d\nu}{2\pi}e^{-i\nu\tau}(A^{\alpha\beta})^{-1}(\nu).
\end{equation}
Here, $(A^{\alpha\beta})^{-1}(\nu)$ is the inverse of the action matrix defined in Eq.~\eqref{eq:amatrix} with $\chi_a=\chi_b=0$. The determinant of the action matrix is given by $p(\nu)$ in Eq.~\eqref{eq:pomega}. In general, $p(\nu)$ has four complex roots $\nu_1,\,\nu_1^\ast,\,\nu_2,\,\nu_2^\ast$ with
\begin{align}
&\nu_1=\frac{1}{2}\left(i-\sqrt{-r-2i\gamma\Delta}\right),\\
&\nu_2=\frac{1}{2}\left(i+\sqrt{-r-2i\gamma\Delta}\right).
\end{align}

In order to perform the inverse Fourier transform, we have to separate the advanced ($\tau>0$) and retarded ($\tau<0$) parts of the propagator. Then, we can perform the integral by closing the contour in the lower (upper) halve of the complex plane. For the advanced propagator $G^A$ we obtain
\begin{equation}
G^A(\tau)=2\pi G_Q\left[ G_1^A e^{-i\nu_1^*\tau}+G_2^Ae^{-i\nu_2^*\tau}\right],
\end{equation}
with 
\begin{align}
&G_1^A=\frac{B(\nu_1^*)}{ir+2\gamma\Delta+\sqrt{-r+2i\gamma\Delta}(1-i\gamma\Delta)},\\
&G_2^A=\frac{B(\nu_2^*)}{ir+2\gamma\Delta-\sqrt{-r+2i\gamma\Delta}(1-i\gamma\Delta)}.
\end{align}
Analogously, we obtain for the retarded propagator $G^R$ 
\begin{equation}
G^R(\tau)=2\pi G_Q\left[ G_1^R e^{-i\nu_1\tau}+G_2^R e^{-i\nu_2\tau}\right],
\end{equation}
with 
\begin{align}
&G_1^R=\frac{B(\nu_1)}{ir-2\gamma\Delta-\sqrt{-r-2i\gamma\Delta}(1+i\gamma\Delta)},\\
&G_2^R=\frac{B(\nu_2)}{ir-2\gamma\Delta+\sqrt{-r-2i\gamma\Delta}(1+i\gamma\Delta)}.
\end{align}
Here, the $4\,\times \,4$ matrix $B(\nu)$ is given by
\begin{widetext}
\begin{align}
B(\nu)=\ &n_a\begin{bmatrix}
f_{21}^a(\nu,\gamma) &f_{14}(-\nu,-\gamma)  \\
-f_{14}(-\nu^*,-\gamma)^* &f_{12}^b(-\gamma) \end{bmatrix}\otimes \begin{bmatrix}
1 \!& 1\\
1 \!& 1
\end{bmatrix}+n_b\begin{bmatrix}
f_{12}^a(\gamma)  &f_{14}(\nu,\gamma)  \\
-f_{14}(\nu^*,\gamma)^* &f_{21}^b(-\nu,-\gamma) \end{bmatrix}\otimes \begin{bmatrix}
1 \!& 1\\
1 \!& 1
\end{bmatrix}\\
&+\begin{bmatrix}
f_{11}^a(\nu,\gamma) & f_{12}^a(\gamma) &f_{13}(\nu,\gamma) &f_{14}(\nu,\gamma) \\
f_{21}^a(\nu,\gamma) &-f_{11}^a(\nu^*,\gamma)^* &f_{14}(-\nu,-\gamma) &f_{24}(\nu,\gamma) \\
-f_{24}(\nu^*,\gamma)^* &-f_{14}(\nu^*,\gamma)^* &f_{11}^b(-\nu,-\gamma) &f_{21}^b(-\nu,-\gamma) \\
-f_{14}(-\nu^*,-\gamma)^*&-f_{13}(\nu^*,\gamma)^* &f_{12}^b(-\gamma)&-f_{11}^b(-\nu^*,-\gamma)^*\end{bmatrix},
\end{align}
\end{widetext}
where we use the abbreviations
\begin{align}
&g(\nu,\gamma)=i-i\gamma+\Delta-2\nu,\nonumber\\
&h(\nu,\gamma)=(i-\Delta)^2-\epsilon^2+\gamma^2(1+\epsilon^2)+4i\gamma\nu-4\nu^2,\nonumber\\
&f_{11}^k(\nu,\gamma)=Z_k^0 g(\nu,\gamma)h(\nu,\gamma),\nonumber\\
&f_{12}^k(\gamma)=-2iZ_k^0(1-\gamma)^2(1+\gamma) \epsilon^2,\nonumber\\
&f_{21}^k(\nu,\gamma)=-2iZ_k^0(1+\gamma)g(\nu,\gamma)g(\nu^*,\gamma)^*,\nonumber\\
&f_{13}(\nu,\gamma)=\sqrt{Z_a^0Z_b^0}\sqrt{1-\gamma^2} h(\nu,\gamma)\epsilon,\nonumber\\
&f_{14}(\nu,\gamma)=-2i\sqrt{Z_a^0Z_b^0}\sqrt{1-\gamma^2}(1-\gamma) g(-\nu^*,-\gamma)^*\epsilon,\nonumber\\
&f_{24}(\nu,\gamma)=-\sqrt{Z_a^0Z_b^0}\sqrt{1-\gamma^2} [h(\nu^*,\gamma)^*+4(1-\gamma^2)]\epsilon.\nonumber
\end{align}
\section{FCS integral}\label{app:integral}
We show how one can calculate integrals of the type
\begin{equation}
 I(\zeta) = -\int\frac{d\omega}{2\pi} \ln\left(1-\frac{\zeta}{g(\omega)}\right)
\end{equation}
with $g(\omega)$ a polynomial with real coefficients and $g(\omega) \propto \omega^{2n}$, $n\in\mathbb{N}$ for $\omega \to \infty$ such that the integral converges. In the following, we consider $\zeta
\in \mathbb{R}$. However, the results can be analytically continued to complex
$\zeta$.

The first step is to integrate by parts with the result
\begin{equation}\label{eq:integral}
 I(\zeta) = \int\frac{d\omega}{2\pi} \frac{\zeta \omega g'(\omega)}{g(\omega)
 [g(\omega ) - \zeta]}.
\end{equation}
We denote with $a_1, \cdots, a_n, a_1^*, \cdots, a_n^*$ the (simple) roots of
the polynomial $g(\omega)$ that are chosen such that $\operatorname{Im} a_j
>0$. Similar, $b_1(\zeta), \cdots, b_n(\zeta), b_1^*(\zeta), \cdots,
b_n^*(\zeta)$ denote the roots of the polynomial $g(\omega) -\zeta$ which coincide
with $a_j$ for $\zeta =0$.

The integral \eqref{eq:integral} can then be evaluated with the residue
theorem. In particular, we obtain
\begin{equation}\label{eq:integralsol}
 I(\zeta) = i\sum_{j=1}^n b_j(\zeta) - i \sum_{j=1}^n a_j \;.
\end{equation}
For $n=2$, we can use the following property of a depressed quartic polynomial to calculate the sum: given a
polynomial of the form
\begin{equation}
 g(\omega) = \omega^4 + p \omega^2 + q \omega +r(\zeta),
\end{equation}
the negative square of sum of the roots in the upper half plane, $m(\zeta) =
-[b_1(\zeta)+b_2(\zeta)]^{2}$, is a root of the resolvent cubic
\begin{equation}
  m^3 - 2 p m^2 +[p^2 -4 r(\zeta)] m +q^2\;.
\end{equation}
With this property, Eq.~\eqref{eq:integralsol} reduces to
\begin{equation}
 I(\zeta) = \sqrt{m(\zeta=0)} -\sqrt{m(\zeta)}\;.
\end{equation}
\section{Full solution of the cumulant-generating function}\label{app:fullm}
The cumulant-generating function is given by Eq.~\eqref{eq:lambdam} with $m$ the solution of Eq.~\eqref{eq:cubic} that coincides with $1$ for $s=0$. We use the trigonometric method to calculate an analytic expression for the relevant solution of $m$ and obtain
\begin{widetext}
\begin{align}
m=\frac{1}{3}\left\lbrace 1+r+2\sqrt{1-r+r^2-3\tilde s+3\gamma^2\Delta^2}\cos\left[\frac{1}{3}\arccos \left(\frac{(1+r)(2-5r+2r^2-9\tilde s)+9\gamma^2\Delta^2(r-2)}{2(1-r+r^2-3\tilde s+3\gamma^2\Delta^2)^{3/2}}\right)\right] \right\rbrace,
\end{align}
\end{widetext}
with $\tilde s=(1-\gamma^2)^2 \epsilon^2 s$. However, in many situations, e.g. the calculation of the cumulants, the implicit solution of $m$ in Eq.~\eqref{eq:cubic} is considerably easier to use.

\section{Non-analyticity of cumulant-generating function}\label{app:analytic}

For the extreme statistics in Sec.~\ref{sec:bursts}, we need to find the minimum value $\mu_0>0$ of $\ln[1+1/\rho(\nu)]$ as a function of frequency $\nu$. We assume a perfect counting efficiency, $f_a=f_b=1$. Then, Eq.~\eqref{eq:schi} is reduced to $s(\chi)=e^{i\chi} -1$ with $\chi=\chi_a+\chi_b$ and the cumulant-generating function in Eq.~\eqref{eq:lambdaint} is given by
\begin{align}
\lambda(\chi)=- \tau \int\frac{d\nu}{2\pi}
 \ln\left[1 - \rho(\nu)(e^{i\chi} -1)\right].
\end{align}
This cumulant-generating function has a singularity at $\chi=-i\mu_0$. Therefore, we can use our result for $\lambda(\chi)$ in Eq.~\eqref{eq:lambdam} to learn more about $\mu_0=\ln (s_0+1)$. From Eq.~\eqref{eq:cubic}, it is clear that $m$ can only coincide with $0$ for $\gamma\Delta =0$. Therefore, the square root in Eq.~\eqref{eq:lambdam} can only cause the non-analyticity in this special case. However, as the solution for $s_0$ needs to be continuous, it is sufficient to find the point of non-analyticity of the relevant solution $m$.
We take the discriminant of Eq.~\eqref{eq:cubic} and obtain a cubic equation for $\tilde s=(1-\gamma^2)^2 \epsilon^2 s$
\begin{align}\label{eq:sstar}
&(r\!-\!1\!\!+\!\gamma^2\!\Delta^{\!2})(r^2\!\!+\!4\gamma^2\!\Delta^{\!2})
\!=\!4\tilde s^3\!\!-\!\tilde s^2(1\!\!-\!10r\!+\!r^2\!\!+\!12\gamma^2\!\Delta^{\!2})\nonumber\\
&-\!2\tilde s[r^3-4r^2+r-\gamma^2\Delta^2(r^2-r+10+6\gamma^2\Delta^2)].\end{align}
The relevant point of non-analyticity of the solution $m$ is given by the solution of Eq.~\eqref{eq:sstar} that coincides with $0$ at the threshold where $r=1-\gamma^2\Delta^2$. In the limits previously discussed, we can express the result analytically.
\subsection{No detuning or no asymmetry}
For $\gamma\Delta=0$, the solution has a non-analytic point (second-order derivative is discontinuous) at $r=-1$. The solution is given by
\begin{equation}
 \mu_0= \begin{cases}
\ln \left[1-\frac{r}{(1-\gamma^2)^2 \epsilon^2}\right] & \mathrm{if} \,\,\, r < -1 \\
\ln \left[1+\frac{(r-1)^2}{4(1-\gamma^2)^2 \epsilon^2}\right] & \mathrm{if} \,\,\, r > -1,
\end{cases}
\end{equation}
which qualitatively corresponds to the solution for the degenerate oscillator\cite{Padurariu:12}. For $\gamma\Delta\neq 0$, the discontinuity of the second-order derivative is washed out. This is a qualitative difference between the non-degenerate and the degenerate oscillator.

\subsection{Close to threshold}
In the vicinity of the threshold, we obtain
\begin{align}
\mu_0=\frac{(\epsilon_*-\epsilon)^2}{1+\gamma^2\Delta^2},
\end{align}
which corresponds to the $\mu_0$ obtained in Eq.~\eqref{eq:mu0}.


\begin{thebibliography}{10}
\bibitem{commut}
To avoid mutual excitations between both resonators, the ratio
  $\Omega_a/\Omega_b$ should be hard to approximate by the ratio of two small
  integer numbers. This is provided, e.g., by the golden ratio.
\bibitem{chi}
Note that this modification is done in time space with the piecewise-constant
  counting fields $\chi_k(t)$.
\bibitem{fano}
Note that the summand $1$ is missing in the definition of $F_{ab}$ compared to
  the definition of $F_{kk}$ since there is no 'self'-correlation.
\end{thebibliography}
\end{document}